# High-field magneto-optical imaging of superconducting critical states beyond 10 T using a paramagnetic garnet sensor

Yuto Kinoshita[1,*], Masayuki Toyoda[2], Yoshiaki Kobayashi[2], Masayuki Itoh[2], and Masashi Tokunaga[1]

[1]*Institute for Solid State Physics, The University of Tokyo, Kashiwa 277-8581, Japan*

[2]*Department of Physics, Nagoya University, Nagoya 464-8601, Japan*

* kinoshita@issp.u-tokyo.ac.jp

Spatially resolved characterization of the critical current density $J_c$ in superconductors under high magnetic fields is crucial for both fundamental understanding and practical applications. However, conventional techniques primarily provide bulk-averaged values, making it difficult to resolve local variations of $J_c$, especially in high magnetic fields. In this work, we develop a magneto-optical imaging (MOI) technique that enables visualization of superconducting critical states in steady magnetic fields up to 13 T. This is achieved by employing a paramagnetic Nd-garnet indicator combined with a polarizing microscope system. Using this method, we directly image the magnetic flux distribution in a bulk single crystal of an iron-based superconductor Ba(Fe$_{1-x}$Co$_x$)$_2$As$_2$ ($x$ = 0.075) at 12 K and 20 K across the entire sample area (~1 mm). From the measured magnetic field distributions, we quantitatively reconstruct the spatial distribution of the critical current density. The extracted field dependence of $J_c$ is in good agreement with that obtained from conventional magnetization measurements. Furthermore, we demonstrate vector mapping of current flow within the sample by converting the magnetic field distribution into local current-density distributions. Our results establish high-field MOI as a powerful approach for spatially resolved evaluation of superconducting critical states and provide a new pathway for investigating inhomogeneous current transport in superconductors under high magnetic fields.

# I. INTRODUCTION

Superconductors are widely used in high magnetic field applications such as magnets, power devices, and energy systems, where the critical current density $J_c$ is a key parameter that determines performance [1,2]. In practical materials, $J_c$ is often strongly influenced by spatial inhomogeneity arising from defects, pinning centers, and microstructural variations [1,2]. Therefore, spatially resolved characterization of $J_c$ is essential for understanding current transport properties and optimizing superconducting materials. Magneto-optical imaging (MOI) is a powerful technique for visualizing magnetic flux distributions in superconductors with high spatial resolution, enabling direct observation of critical states and current flow [3-5]. By analyzing the magnetic field distribution, the local current density can be reconstructed and $J_c$ can be evaluated with spatial resolution. This approach has been widely applied to investigate flux penetration, current distribution, and pinning properties in various superconducting systems [3]. In this method, a magneto-optical indicator film, typically a ferrimagnetic garnet, is placed on the sample surface, and the stray magnetic field from the sample is detected via the Faraday rotation of the indicator, which is converted into intensity variations through a polarizer–analyzer optical setup. MOI using ferrimagnetic garnet indicators has been widely applied, for example, to YBCO thin films, and has provided important insights into flux penetration and critical states [6]. However, the magnetization of ferrimagnetic garnets saturates at magnetic fields below approximately 1 T, which fundamentally limits their applicability in high magnetic fields. Alternative approaches using paramagnetic indicators of europium chalcogenides such as EuSe have also been explored [7,8], but the useful field range is limited to ~3 T and is therefore not suitable for measurements in magnetic fields exceeding 10 T. As a result, conventional MOI techniques are fundamentally limited to relatively low magnetic fields, typically below a few teslas, and spatially resolved measurements of superconducting critical states under high magnetic fields—where many technologically relevant phenomena emerge—have remained largely inaccessible. To overcome this limitation, alternative approaches are required to extend MOI to higher magnetic fields. Paramagnetic garnet materials have been reported to exhibit large magneto-optical effects without saturation even under magnetic fields of several tens of tesla, owing to crystal-field interactions. Representative examples include $Nd_3Ga_5O_{12}$ (NdGG) and $Yb_3Ga_5O_{12}$ (YbGG) [9,10]. These properties suggest that paramagnetic garnets are promising candidates for magneto-optical indicators in high magnetic fields. In this work, we develop a high-field MOI technique based on a paramagnetic Nd-garnet indicator integrated with a polarizing microscope system, enabling visualization of superconducting critical states in steady magnetic fields up to 13 T. Using this method, we investigate a bulk single crystal of the iron-based superconductor $Ba(Fe_{1-x}Co_x)_2As_2$ ($x = 0.075$) and directly image magnetic flux distributions over the entire sample area at low temperatures. $Ba(Fe_{1-x}Co_x)_2As_2$ is a representative iron-based superconductor with relatively low anisotropy and high upper critical fields, making it an ideal platform for studying vortex dynamics and current transport under high magnetic fields [11-13]. While its critical current density and vortex pinning properties have been extensively investigated using bulk techniques such as magnetization and transport measurements [11,12], spatially resolved characterization of the critical current density—particularly in high magnetic fields—remains largely unexplored. Although MOI of remanent flux distributions has been reported [11-13], its extension to high magnetic fields has remained challenging.

Here, we quantitatively reconstruct the spatial distribution of the critical current density from measured magnetic field profiles and further demonstrate vector mapping of the current flow within the sample. These results establish high-field MOI as a powerful tool for spatially resolved characterization of superconducting properties under high magnetic fields and open new opportunities for investigating inhomogeneous current transport in superconductors.

**II. EXPERIMENTAL SETUP**

The experimental setup of the MOI system is shown in Fig. 1(a). A superconducting magnet system (Physical Property Measurement System, Quantum Design) was used to apply magnetic fields [14]. The sample was mounted on a standard sample puck and positioned at the center of the magnetic field. The optical system was constructed on an aluminum frame equipped with an *x–y* translation stage. A plano-convex lens (LA1464-A, Thorlabs) with a focal length of 1000 mm was used for imaging and mounted on a fiber-reinforced plastic (FRP) pipe. The pipe was inserted into the PPMS sample chamber through a Wilson seal, allowing optical access to the sample space. The focus was adjusted by vertically translating the FRP pipe. The commercial polarizing microscope (BMFX, Olympus) was used. The incident light was provided in a coaxial reflection geometry using a half mirror. Light from an LED source (M455L4 and M625L4, Thorlabs) was linearly polarized and directed onto the sample. The reflected light was collected and imaged onto a CCD camera (ML695, Finger Lakes Instrumentation) through a 5× TV lens. An analyzer with an adjustable angle was placed in the detection path to measure polarization rotation.

A commercially available $Nd_3Ga_5O_{12}$ (NdGG) single crystal with (001) surface orientation (10 mm × 10 mm × 0.5 mm) was used as the magneto-optical indicator. For handling and mechanical stability, the NdGG crystal was bonded to a 0.5-mm-thick quartz substrate using atomic diffusion bonding via a 5-nm-thick $Y_2O_3$ intermediate layer [15,16]. The NdGG crystal was then polished down to a thickness of 10 μm. Finally, an Al reflective layer was deposited on the NdGG surface by vacuum evaporation. A schematic illustration of the measurement principle is shown in Fig. 1(b). The sample was attached to the Al-coated side of the indicator, and imaging was performed from the quartz substrate side. Incident light is reflected at the Al layer and propagates twice through the NdGG layer. The Faraday rotation accumulated during this propagation depends on the total magnetic field, including both the applied magnetic field and the stray field from the sample. Therefore, the spatial distribution of the magnetic field can be obtained by detecting the polarization rotation of the reflected light.

A (001)-oriented bulk single crystal of Ba(Fe$_{1-x}$Co$_x$)$_2$As$_2$ ($x$ = 0.075) grown by the Bridgman method was used as the sample. The sample size was approximately 1.1 mm × 1.3 mm × 0.26 mm. The sample was attached to the indicator using Apiezon N grease (M&I Materials Ltd), and the indicator was mounted on the sample puck using the same grease. Magnetization measurements were performed on the same sample using a Magnetic Property Measurement System (MPMS, Quantum Design), with magnetic fields up to 7 T.

## III. RESULTS AND DISCUSSION

**A. Magnetic-field dependence of Faraday rotation in the magneto-optical indicator**

The magnetic-field dependence of the Faraday rotation of the indicator was characterized prior to imaging measurements. The measurements were performed using only the magneto-optical sensor without the sample. LEDs with wavelengths of 625 nm and 455 nm were used as light sources. The measurements were conducted at 20 K for both wavelengths and at 12 K for 455 nm up to high magnetic fields. At each magnetic field, a series of images was acquired while rotating the analyzer angle $\theta$. The intensity averaged over the central region of the sensor was fitted using the following equation,

$$I = I_0 \sin^2(\theta - \theta_0) + I_{\min}, \quad (1)$$

where $I_0$ is the incident light intensity, $\theta_0$ is the Faraday rotation angle, and $I_{\min}$ represents the background signal. The extracted $\theta_0$ as a function of magnetic field is shown in Fig. 1(c). At 20 K, the rotation angle increases monotonically with magnetic field up to 12 T for both wavelengths [open red squares (625 nm) and open blue squares (455 nm)]. The rotation at 455 nm is approximately three times larger than that at 625 nm. At 12 K and 455 nm (open blue circles), the rotation also increases monotonically, although a slight tendency toward saturation is observed at high magnetic fields. Importantly, across all measurement conditions, the Faraday rotation does not saturate and remains a monotonic function of the magnetic field, demonstrating that the paramagnetic indicator is suitable for high-field MOI. When considering a field range of ±1 T around each magnetic field value, the Faraday rotation angle can be expressed as $\theta_0 + \Delta\theta_0$, where $\theta_0$ is the rotation angle at the reference field and $\Delta\theta_0$ is the field-induced change. In this range, $\Delta\theta_0$ remains small (less than ~5°), and its dependence on the magnetic field can be regarded as approximately linear, as shown in Fig.1(c). Under this condition, Eq. (1) can be expanded in $\Delta\theta_0$. Since $\Delta\theta_0$ is sufficiently small, the higher-order terms can be neglected, and the intensity $I$ can be approximated as a linear function of $\Delta\theta_0$ with a constant offset. Consequently, within the ±1 T range, the intensity $I$ is approximately proportional to the magnetic field. At 20 K and 10 T, the rotation per unit length is estimated to be 5000 deg/cm (455 nm) and 2000 deg/cm (625 nm), which are comparable with reported values of 4000 deg/cm (490 nm) and 2000 deg/cm (610 nm) in similar wavelength ranges [9]. Based on these results, the 455 nm LED was used for subsequent imaging experiments due to its larger Faraday rotation.

## B. High-field magneto-optical imaging of superconducting critical states in bulk $Ba(Fe_{1-x}Co_x)_2As_2$

Figure 2(a) shows an optical image of the sample. Figure 2(b) presents a reference image of the magneto-optical indicator taken at 35 K and 0 T without the sample contribution. Dark spots and lines correspond to defects and scratches on the indicator surface. The sample is located beneath this region of the indicator. The sample was zero-field cooled to 12 K and subsequently subjected to a magnetic field of 1 T. The resulting image is shown in Fig. 2(d). A dark region corresponding to the sample shape is observed near the center, indicating a reduced magnetic field due to the diamagnetic response of the superconductor. After increasing the field to 2 T and then reducing it back to 1 T, the image shown in Fig. 2(e) was obtained. In contrast to Fig. 2(d), a bright contrast appears inside the sample, corresponding to trapped magnetic flux induced during the field reduction. The magnetization curve measured for the same sample is shown in Fig. 2(c). The red curve represents the major loop measured between −7 T and 7 T at 12 K, while the blue curve shows a minor loop between 0 and 2 T. The agreement between the minor and major loops indicates that full flux penetration is achieved within this field range. The amount of trapped flux corresponds to the magnetization difference $\Delta M$ at 1 T. Therefore, the spatial distribution of the corresponding magnetic flux can be obtained by taking the difference between the images in Figs. 2(d) and 2(e). The resulting differential image is shown in Fig. 2(f), where a clear contrast is observed despite residual background originating from indicator imperfections. The intensity increases monotonically from the sample edge toward the center, and a flux penetration pattern reflecting the sample geometry is evident.

To obtain quantitative magnetic field distributions, the image intensity was converted into magnetic flux density using a calibration procedure. The contrast between the up-and down-sweep images is symmetric, and their sum yields a background image corresponding to the external magnetic field (the sum image), free of the sample stray-field contribution. Assuming a linear relation between intensity and magnetic field in the range of 0–2 T, the difference between the sum image at 2 T and the zero-field image was used to determine the intensity change per unit field. The differential image [Fig. 2(f)] was then normalized using this factor to obtain the magnetic flux density distribution shown as Fig. 3(d). This procedure effectively removes background contributions from the indicator. Using the same method, flux density distributions were obtained at 4, 6, 10, and 13 T from ±1 T field sweeps, as shown in Figs. 3(e)–3(h). Corresponding results at 20 K for 1, 4, and 6 T are shown in Figs. 3(a)–3(c), where the measurements were performed with ±0.5 T field sweeps. The color scale is shown in Fig. 3. At 12 K, the overall flux distribution pattern remains similar to that at 1 T, while the contrast increases with magnetic field up to 6 T and then decreases at higher magnetic fields (10 T and 13 T). In contrast, at 20 K, the flux contrast is significantly reduced: at 1 T the contrast is already smaller than that at 12 K, and it nearly disappears at 4 T and vanishes at 6 T. These results demonstrate that the temperature dependence of the trapped flux and its non-monotonic field dependence at 12 K directly reflect the second magnetization peak (SMP), in agreement with previous studies [17-18]. The trapped magnetic flux reflects the critical current density $J_c$. To quantitatively evaluate $J_c$, we analyzed the line profiles of the magnetic flux density distributions shown in Figs. 3(a)–3(h).

The sample used in this study is a bulk slab geometry, and its critical state can be described by the Bean model [19]. Assuming that $J_c$ is field-independent locally, the out-of-plane magnetic flux density $B_z$ satisfies

$$\frac{dB_z}{dx} = \mu_0 J_c, \quad (2)$$

where $x$ is the in-plane coordinate and $\mu_0$ is the permeability of free space, $4\pi \times 10^{-7}$ Tm/A. Therefore, $J_c$ can be obtained from the slope of the magnetic flux density profile. Figures 4(a) and 4(b) show representative line profiles at 12 K and 20 K, respectively. The insets show the magnetic flux distribution at 1 T, with red lines indicating the positions used for the line-profile analysis. At both temperatures, the magnetic flux density varies approximately linearly from the sample edge toward the center, indicating that the critical state is well described by the Bean model. To estimate $J_c$, the linear regions on both sides of the sample center were fitted with straight lines, and the average slope was used to calculate $J_c$. The resulting values, denoted as $J_c^{MOI}$, are plotted in Fig. 4(e) (open red circles for 12 K and open red squares for 20 K). Similar analyses were performed at additional magnetic fields not shown in Fig. 3 and the obtained $J_c^{MOI}$ values are included in Fig. 4(e) (see Supplementary Material). At 12 K, $J_c^{MOI}$ increases with magnetic field up to 8 T and then decreases toward 13 T. At 20 K, $J_c^{MOI}$ exhibits a maximum around 2 T and decreases with increasing field, approaching zero near 6 T, indicating a transition to the normal state.

For comparison, $J_c$ was also evaluated from magnetization measurements performed on the same sample using the same field sweep protocol. The results are shown in Figs. 4(c) and 4(d) for 12 K and 20 K, respectively. The blue curves represent minor loops (±1 T at 12 K and ±0.5 T at 20 K), while the red curves correspond to the major loop (−7 T to 7 T). The agreement between minor and major loops indicates full flux penetration. The critical current density $J_c^{MH}$ was calculated from the magnetization hysteresis width $\Delta M$ using the extended Bean model for a rectangular slab with dimensions $2a \times 2b$ ($a < b$) [20]:

$$J_c = \frac{\Delta M}{a(1-\frac{a}{3b})}. \quad (3)$$

The resulting $J_c^{MH}$ values are plotted in Fig. 4(e) (open blue circles for 12 K and open blue squares for 20 K). The same procedure was applied to magnetization curves not shown in Fig. 4(c) and 4(d), and the corresponding $J_c^{MH}$ values are also included in Fig. 4(e) (see Supplementary Material). Here, a = 0.55 mm and b = 0.65 mm were used. Although the absolute values of $J_c^{MOI}$ are approximately three times smaller than those of $J_c^{MH}$, both methods show consistent temperature scaling and similar qualitative field dependence. This agreement indicates that the present MOI-based method provides a reliable estimate of $J_c$. The discrepancy in absolute values is likely due to magnetic field attenuation caused by the finite distance between the sample and the indicator. Nevertheless, the ability to spatially resolve the critical state and quantitatively estimate $J_c$ under high magnetic fields up to 13 T represents a significant advancement, as such measurements have not been previously reported.

## C. Vector mapping of critical current density from magnetic flux density distributions

To fully exploit the spatially resolved critical-state information obtained by the present method, we further attempted to reconstruct the vector distribution of the current density from the measured magnetic field distribution. The magnetic field **B**(**r**) at position **r** generated by a current density **J**(**r**′) at position **r**′ is given by the Biot–Savart law as

$$\mathbf{B}(\mathbf{r}) = \frac{\mu_0}{4\pi} \int \frac{\mathbf{J}(\mathbf{r}')\times(\mathbf{r}-\mathbf{r}')}{|\mathbf{r}-\mathbf{r}'|^3} d^3\mathbf{r}'. \quad (4)$$

According to Ref. [21], once the magnetic field distribution is known, the corresponding current density distribution **J**(**r**′) can be reconstructed by performing the calculation in reciprocal space using the Fourier transformation. Details of the reconstruction procedure are provided in the Supplementary Material.

Figure 5(a) shows the magnetic flux density distribution obtained at 12 K and 12 T. This distribution was converted into the current density distribution using the above procedure. The spatial distributions of the $x$- and $y$-components of the current density, $J_x$ and $J_y$, are shown in Figs. 5(b) and 5(c), respectively, and the magnitude $J = |\mathbf{J}|$ is shown in Fig. 5(d). The coordinate system is defined in the lower-left corner of Fig. 5(d). Based on $J_x$ and $J_y$, the current density vectors at each position were mapped and are indicated by red arrows in Fig. 5(d). For clarity, the vectors are subsampled every 20 pixels in both the $x$ and $y$ directions. The direction and length of each arrow represent the direction and magnitude of the local current density, respectively. As seen from the spatial distribution of $J$, the current density is nearly uniform over a wide region and forms circulating patterns, consistent with the critical-state behavior described by the Bean model. In contrast, a region near the sample center where the current density is significantly reduced is clearly observed. According to theoretical predictions for finite-thickness slab samples, a current-free region can appear in the central part of the sample in the critical state [22]. The present result is consistent with this prediction. By averaging $J$ over regions where the current density is approximately uniform, the current density at 12 T is estimated to be on the order of $1 \times 10^9$ A/m$^2$, which is approximately twice the value obtained from the line-profile analysis and about 70% of $J_c^{\mathrm{MH}}$. This improvement is attributed to partial compensation for magnetic field attenuation arising from the finite distance between the sample and the indicator, which is accounted for in the reconstruction process. To the best of our knowledge, this is the first demonstration of vector-resolved critical current density mapping in a bulk superconductor under magnetic fields exceeding 10 T. This method provides a powerful tool for identifying spatially inhomogeneous current-limiting regions in superconductors.

## IV. SUMMARY

In this work, we developed a magneto-optical imaging (MOI) technique capable of visualizing superconducting critical states under high magnetic fields up to 13 T by employing a paramagnetic Nd-garnet indicator. This approach overcomes the intrinsic limitation of conventional MOI, arising from saturation of the indicator magnetization, and enables imaging in a previously inaccessible high magnetic field regime. Using this technique, we successfully obtained spatially resolved magnetic flux density distributions in a bulk Ba(Fe$_{1-x}$Co$_x$)$_2$As$_2$ single crystal at 12 K and 20 K. The extracted flux profiles exhibit behavior consistent with the critical state described by the Bean model. From these distributions, the critical current density $J_c$ was quantitatively evaluated using line-profile analysis, and the resulting field dependence agrees well with that obtained from conventional magnetization measurements. Furthermore, we demonstrated the reconstruction and vector mapping of the current density from the magnetic flux density distribution. The resulting current maps reveal circulating current patterns and a current-free region near the sample center, consistent with theoretical predictions for finite-thickness samples. The reconstructed current density reaches values on the order of $10^9$ A/m$^2$ at 12 T, approaching those obtained from magnetization measurements.

To the best of our knowledge, this is the first demonstration of spatially resolved critical current density mapping and vector-resolved current flow in a bulk superconductor under magnetic fields exceeding 10 T using MOI. The present method provides a powerful tool for investigating spatially inhomogeneous current transport and identifying current-limiting regions in superconductors under high magnetic fields.

**Acknowledgements**

We thank Takehito Shimatsu and Miyuki Uomoto (FRIS, Tohoku University) for bonding the NdGG to a glass substrate using the atomic diffusion bonding method. This work was supported by MEXT KAKENHI Grant Number JP23H04862, JSPS KAKENHI Grant Number JP26K17089, and the Daiichi-Sankyo "Habataku" Support Program for the Next Generation of Researchers.

# References

[1] Gurevich A 2011 Nature Mater. **10** 255

[2] Blatter G, Feigel'man M V, Geshkenbein V B, Larkin A I and Vinokur V M 1994 Rev. Mod. Phys. **66** 1125

[3] Jooss C, Albrecht J, Kuhn H, Leonhardt S and Kronmüller H 2002 Rep. Prog. Phys. **65** 651

[4] Ooi S, Tachiki M, Mochiku T, Ito H, Kubo T, Kikuchi A, Arisawa S and Umemori K 2025 Phys. Rev. B **111** 094519

[5] Goa P E, Hauglin H, Olsen Å A F, Baziljevich M and Johansen T H 2003 Rev. Sci. Instrum. **74** 141

[6] Johansen T H, Baziljevich M, Bratsberg H, Galperin Y, Lindelof P E, Shen Y and Vase P 1996 Phys. Rev. B **54** 16264

[7] Schuster Th, Koblischka M R, Ludescher B, Moser N and Kronmüller H 1991 Cryogenics **31** 811

[8] Tokunaga Y, Tokunaga M and Tamegai T 2005 Phys. Rev. B **71** 012408

[9] Guillot M, Zhang F, Xu Y, Yang J H and Wei X 2007 J. Appl. Phys. **101** 09C510

[10] Wang W, Qi X and Liu G 2008 J. Appl. Phys. **103** 073908

[11] Nakajima Y, Taen T and Tamegai T 2009 J. Phys. Soc. Jpn. **78** 023702

[12] Tamegai T, Taen T, Tsuchiya Y, Nakajima Y, Okayasu S and Sasase M 2010 J. Supercond. Nov. Magn. **23** 605

[13] Tamegai T, Tsuchiya Y, Taen T, Nakajima Y, Okayasu S and Sasase M 2010 Physica C **470** S360

[14] Kinoshita Y, Miyakawa T, Xu X and Tokunaga M 2022 Rev. Sci. Instrum. **93** 073702

[15] Shimatsu T and Uomoto M 2010 J. Vac. Sci. Technol. B **28** 706

[16] Shimatsu T, Yoshida H, Uomoto M, Saito T, Moriwaki T, Kato N, Miyamoto Y and Miyamoto K 2021 Proc. 7th Int. Workshop on Low Temperature Bonding for 3D Integration (LTB-3D) 51

[17] H. Yang, H. Luo, Z. Wang, and H.-H. Wen, Appl. Phys. Lett. **93**, 142506 (2008).

[18] B. Shen, P. Cheng, Z. Wang, L. Fang, C. Ren, L. Shan, and H.-H. Wen, Phys. Rev. B **81**, 014503 (2010).

[19] Bean C P 1962 Phys. Rev. Lett. **8** 250

[20] Gyorgy E M, van Dover R B, Jackson K A, Schneemeyer L F and Waszczak J V 1989 Appl. Phys. Lett. **55** 283

[21] Roth B J, Sepulveda N G and Wikswo J P Jr 1989 J. Appl. Phys. **65** 361

[22] Brandt E H 1996 Phys. Rev. B **54** 4246

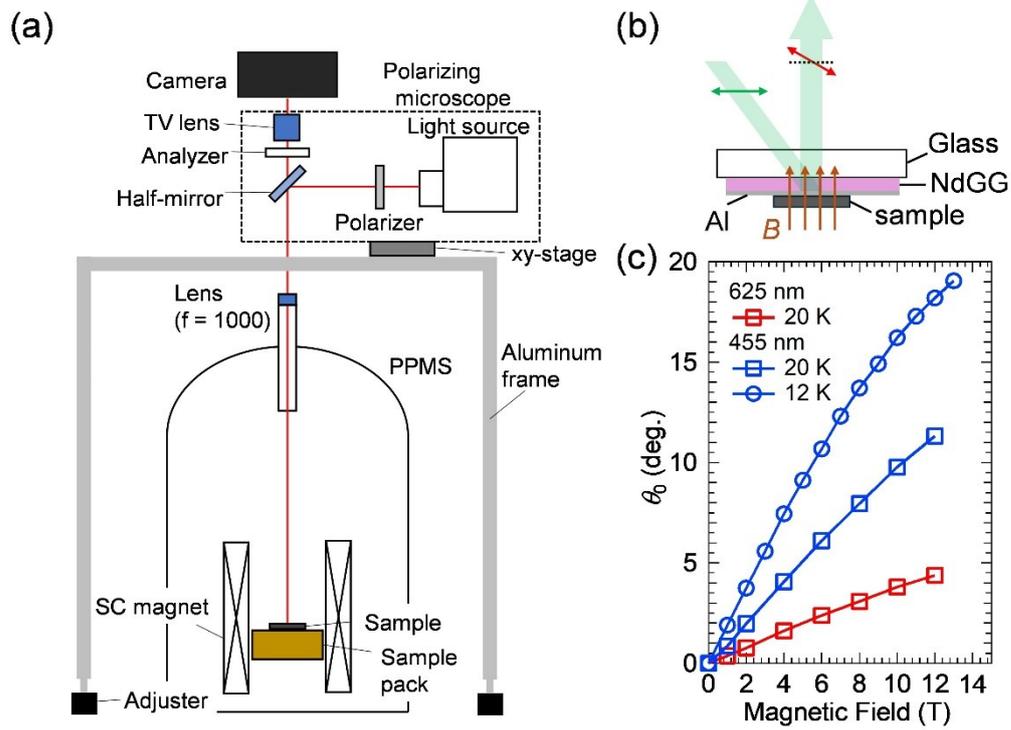

FIG. 1. (a) Schematic of the magneto-optical imaging system under high magnetic fields using a PPMS. (b) Schematic of the magneto-optical sensor. (c) Magnetic-field dependence of the Faraday rotation $\theta_0$ of the sensor. Open red squares represent results at 20 K using a 625 nm LED, while open blue squares and circles correspond to measurements at 20 K and 12 K, respectively, using a 455 nm LED.

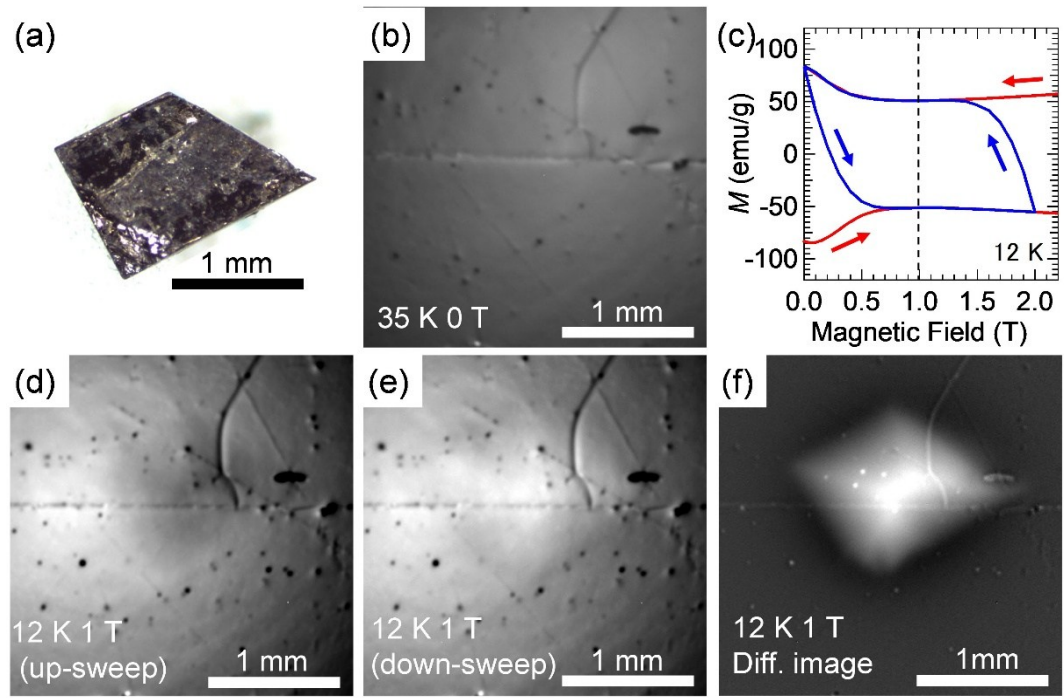

FIG. 2. (a) Optical image of the sample. (b) Microscope image of the magneto-optical sensor at 35 K and 0 T. The sample is located beneath this region of the sensor. (c) Magnetization (*M–H*) curves measured at 12 K. The red curve represents the major loop (−7 to 7 T), and the blue curve represents the minor loop (0 to 2 T). (d) Image at 1 T during the field-increasing (up-sweep) process at 12 K. (e) Image at 1 T during the field-decreasing (down-sweep) process at 12 K. (f) Differential image obtained from (d) and (e).

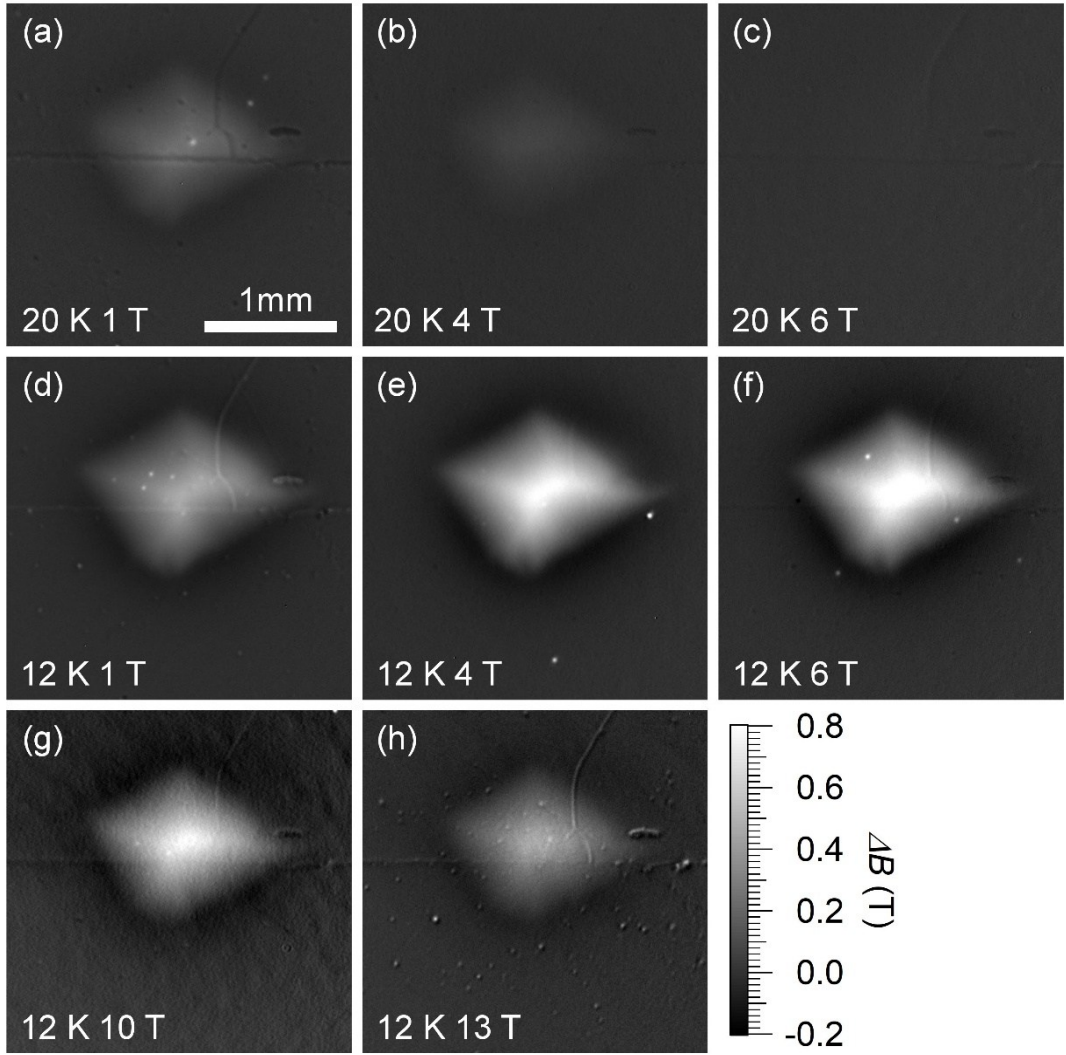

FIG. 3. Differential images of the magnetic flux distribution obtained by subtracting down- and up-sweep images. (a–c) Images at 20 K for magnetic fields of 1, 4, and 6 T, respectively, acquired from field sweeps of ±0.5 T. (d–h) Images at 12 K for magnetic fields of 1, 4, 6, 10, and 13 T, respectively, acquired from field sweeps of ±1 T. The color scale in the lower right represents the magnetic flux density distribution, $\Delta B$.

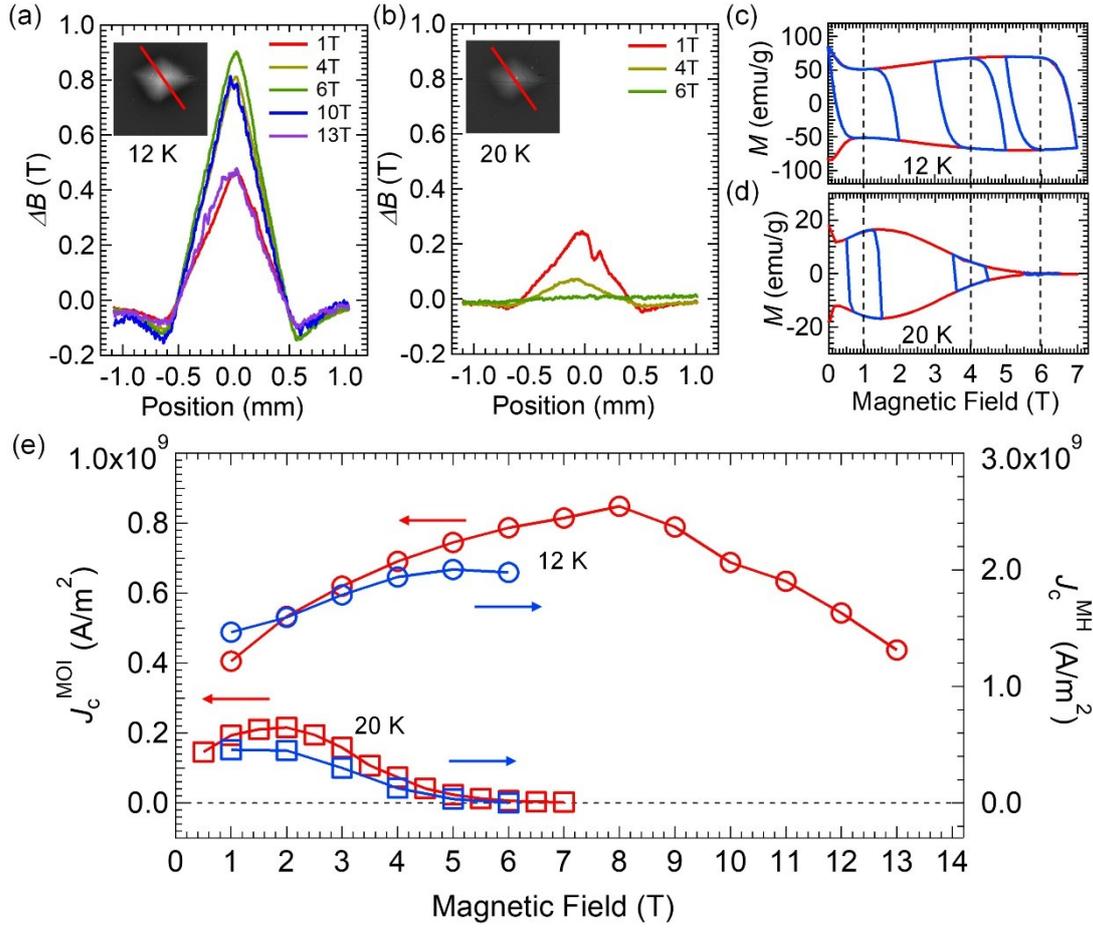

FIG.4. (a,b) Line profiles of $\Delta B$ at various magnetic fields at 12 K and 20 K, respectively. Insets show the $\Delta B$ images at 1 T, where the red lines indicate the positions used for the line-profile analysis. (c,d) Magnetization (*M–H*) curves at 12 K and 20 K, respectively. The red curves represent the major loops (−7 to 7 T), and the blue curves represent the minor loops (±1 T at 12 K and ±0.5 T at 20 K). (e) Magnetic-field dependence of the critical current density $J_c$ at 12 K and 20 K. Open red circles and squares correspond to $J_c^{MOI}$ at 12 K and 20 K, respectively, while open blue circles and squares correspond to $J_c^{MH}$ at 12 K and 20 K, respectively.

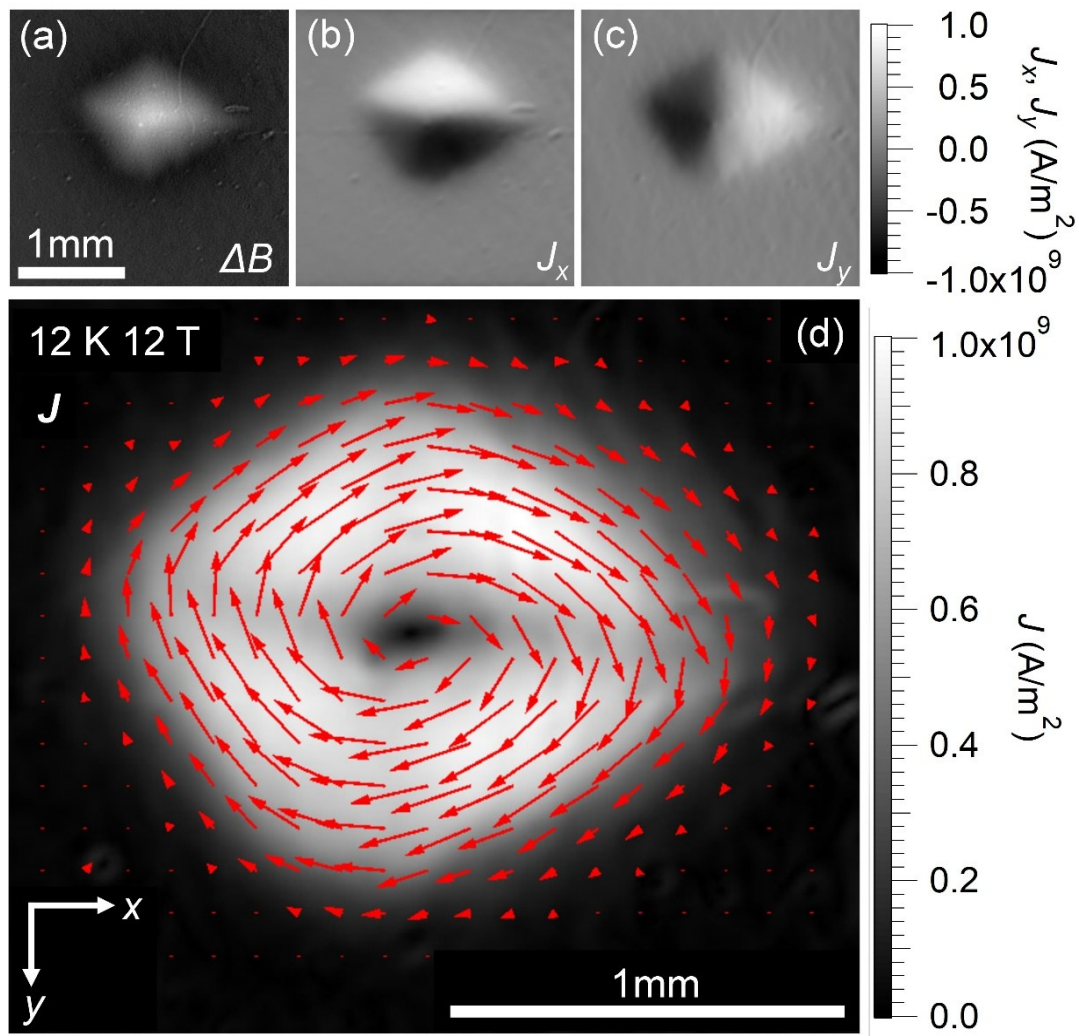

FIG. 5. (a) $\Delta B$ image at 12 K and 12 T. (b,c) Spatial distributions of the current density components $J_x$ and $J_y$, reconstructed from the $\Delta B$ image in (a). (d) Spatial distribution of the current density magnitude $J = |\mathbf{J}|$ and vector mapping of the current density. The coordinate axes are indicated in the lower left.